# High-Intensity Synchrotron Radiation Effects

*Y. Suetsugu*
High Energy Accelerator Research Organization, Tsukuba, Japan

**Abstract**
Various effects of intense synchrotron radiation on the performance of particle accelerators, especially for storage rings, are discussed. Following a brief introduction to synchrotron radiation, the basic concepts of heat load, gas load, electron emission, and the countermeasures against these effects are discussed.

## 1   Introduction

Recent high-power (that is, high-current and high-energy) particle accelerators generate intense synchrotron radiation (SR). This is a good photon source. However, it has the following potentially harmful effects on accelerator performance:

i)   heat load: damage to beam pipes or instruments,

ii)  gas load: short lifetime, noise to particle detectors,

iii) electron emission: beam instabilities, gas load,

iv) radiation: radiation damage.

The first three effects are directly related to the beam and the vacuum system. In this paper, basic and practical concepts to understand the three effects are presented, along with measures to treat these problems, that is, to protect the machine in a broad sense. These problems affect accelerator vacuum systems, but they have widespread effects upon overall machine performances as well. The understanding of these problems is also useful in designing and constructing accelerators.

## 2   Synchrotron radiation

Synchrotron radiation comprises electromagnetic waves emitted when a high-energy charged particle is accelerated in a direction orthogonal to its velocity, such as in a magnetic field (Fig. 1) [1]. The SR is useful as a photon source. The main features of SR compared to other photon sources are:

– high intensity and high photon flux,

– wide range of wavelengths, from infrared to hard X-ray,

– well understood spectrum intensity,

– high brightness,

– high polarization ratio.

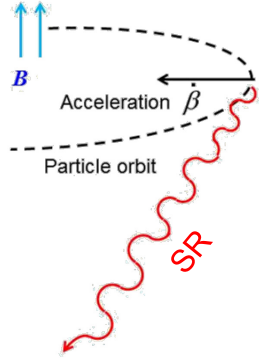

**Fig. 1:** Synchrotron radiation

An accelerated charged particle emits electromagnetic radiation. The radiation fields (electric field $\vec{E}$ and magnetic field $\vec{B}$) are given by using electromagnetic potentials:

$$\vec{E} = -\frac{\partial}{\partial t}\vec{A} - \nabla\phi, \quad \vec{B} = \nabla \times \vec{A} \ . \tag{1}$$

Here, $\phi$ and $\vec{A}$ are the Lienard-Wiechert scalar and vector potentials, which are given by

$$\vec{A}(t) = \frac{e}{4\pi\varepsilon_0 c}\left[\frac{\vec{\beta}}{R(1-\vec{n}\cdot\vec{\beta})}\right]_{ret}, \quad \varphi(t) = \frac{e}{4\pi\varepsilon_0}\left[\frac{1}{R(1-\vec{n}\cdot\vec{\beta})}\right]_{ret}, \tag{2}$$

where $\vec{R}(t_{ret})$ is the distance vector from the source to the observer (see Fig. 2), and $t_{ret}$ is the retarded time, $ct_{ret} = ct - R(t_{ret})$, and $\vec{\beta}$ is the ratio of the velocity $\vec{v}$ to the speed of light $c$ (that is, $\vec{\beta} = \vec{v}/c$). Hence the electric and magnetic fields are obtained by

$$\vec{B} = \frac{1}{c}\left[\vec{n}\times\vec{E}\right]_{ret}, \tag{3}$$

$$\vec{E} = \frac{e}{4\pi\varepsilon_0}\left[\frac{(1-\beta^2)(\vec{n}-\vec{\beta})}{R^2(1-\vec{n}\cdot\vec{\beta})^3}\right]_{ret} + \frac{e}{4\pi\varepsilon_0 c}\left[\frac{\vec{n}\times(\vec{n}-\vec{\beta})\times\dot{\vec{\beta}}}{R(1-\vec{n}\cdot\vec{\beta})^3}\right]_{ret} . \tag{4}$$

At observing points far from the emitting point, the radiation field of the latter term of $\vec{E}$ ($\propto 1/R$) is more important, and the former term can be neglected.

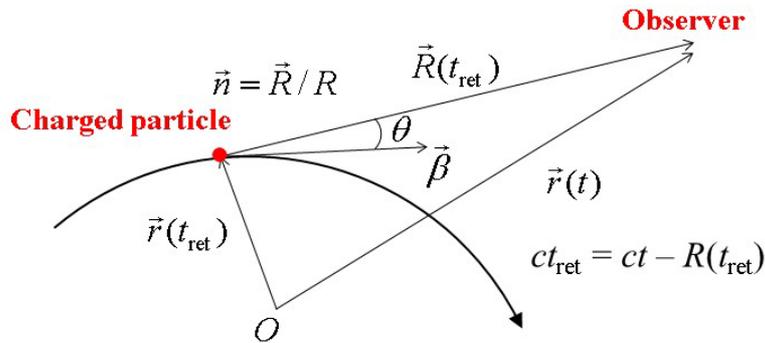

**Fig. 2:** Coordinate system

The pointing vector, that is, the radiation energy flow toward $R$ per unit area, is given by

$$\vec{S}_r(t) = \frac{1}{\mu_0}\vec{E}\times\vec{B} = \frac{1}{\mu_0 c}E^2\left(1-\vec{\beta}\cdot\vec{n}\right)\vec{n}\Big|_{ret} = \varepsilon_0 c E^2\left(1-\vec{\beta}\cdot\vec{n}\right)\vec{n}\Big|_{ret} . \quad (5)$$

Then, the instantaneous differential radiation per unit solid angle d$\Omega$ becomes

$$\frac{dP}{d\Omega} = \vec{n}\cdot\vec{S}R^2\Big|_{ret} = \varepsilon_0 c E^2\left(1-\vec{n}\cdot\vec{\beta}\right)R^2\Big|_{ret} = \frac{e^2}{16\pi^2\varepsilon_0 c}\frac{\left|\vec{n}\times\left\{(\vec{n}-\vec{\beta})\times\dot{\vec{\beta}}\right\}\right|^2}{\left(1-\vec{n}\cdot\vec{\beta}\right)^5}\Bigg|_{ret} . \quad (6)$$

If $\dot{\vec{\beta}}$ is parallel to $\vec{\beta}$,

$$\frac{dP}{d\Omega} = \frac{e^2\dot{\beta}^2}{16\pi^2\varepsilon_0 c}\frac{\sin\theta^2}{(1-\beta\cos\theta)^5} . \quad (7)$$

On the other hand, if $\dot{\vec{\beta}}$ is orthogonal to $\vec{\beta}$,

$$\frac{dP}{d\Omega} = \frac{e^2\dot{\beta}^2}{16\pi^2\varepsilon_0 c}\frac{(1-\beta\cos\theta)^2 - (1-\beta^2)\sin\theta^2}{(1-\beta\cos\theta)^5} . \quad (8)$$

For both cases, when $\beta \approx 1$, the term $(1-\beta\cos\theta)^5$ approaches zero if $\theta$ approaches to zero (see Fig. 3). This means that the power beams to the front of the orbit. This is called 'beaming'. The angle of beaming $\theta$ is given by

$$\theta = \frac{1}{\gamma}, \quad \gamma = \frac{1}{\sqrt{1-\beta^2}} = \frac{E_e}{m_0 c^2} = \frac{E_e[\text{MeV}]}{0.511} \quad \text{(for electron)} . \quad (9)$$

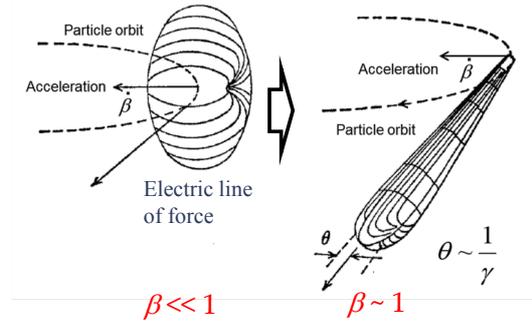

**Fig. 3:** Beaming of SR

Now, consider a charged particle in a homogeneous magnetic field $B$. The acceleration in $B$ is orthogonal to $\beta$, and is given by

$$\dot{\vec{\beta}}_\perp = \frac{\beta^2 c}{\rho} , \quad (10)$$

where the bending radius of charge particle $\rho$ at the electron energy $E_e$ is given by

$$\frac{1}{\rho\,[\text{m}]} = \frac{eBc}{\beta E_e} = 0.2998\frac{B\,[\text{T}]}{\beta E_e\,[\text{GeV}]} . \quad (11)$$

Then, the instantaneous radiation power is obtained by integrating Eq. (8) over $\psi$ (the angle on the plane of $\dot{\vec{\beta}}$) and $\theta$ (the angle orthogonal to the plane of $\dot{\vec{\beta}}$),

$$P = \frac{2cr_e m_e c^2}{3} \frac{\beta^4 \gamma^4}{\rho^2} = \frac{cC_\gamma}{2\pi} \frac{E_e^4}{\rho^2}, \quad C_\gamma \equiv \frac{4\pi}{3} \frac{r_e}{(m_e c^2)^3} = 8.85 \times 10^{-5} \frac{m}{GeV^3}. \quad (12)$$

Here, $r_e$ is the classical electron radius.

Note that the radiation power depends on the mass of the radiating particle as $1/m^4$. Synchrotron radiation is, therefore, much more important for electron and positron rings. For accelerators utilizing a superconducting system, however, such as the LHC, SR is also important for the proton beams, because the heating might transfer a significant heat load to the cryogenics system.

Hereafter, we consider the case of an electron or a positron deflected by a dipole magnet of a ring with a circumference of $C$ (see Fig. 4). The radiation energy along the ring per electron is

$$U_0 = \oint P dt = \frac{C_\gamma}{2\pi} E_e^4 \oint \left( \frac{1}{\rho_x^2} + \frac{1}{\rho_y^2} \right) ds, \quad cdt = ds. \quad (13)$$

For an isomagnetic field (i.e. $\rho$ = constant),

$$U_0 = C_\gamma \frac{E_e^4}{\rho}. \quad (14)$$

For a circulating beam current $I_e$, the total radiation power $P_{Ie}$ becomes

$$P_{Ie} = U_0 \times \frac{I_e}{e} = C_\gamma \frac{E_e^4}{\rho} \times \frac{I_e}{e}, \quad (15)$$

$$P_{Ie} [W] = 8.85 \times 10^4 \frac{E_e [GeV]^4}{\rho [m]} I_e [A] = 2.65 \times 10^4 E_e [GeV]^3 B [T] I_e [A]. \quad (16)$$

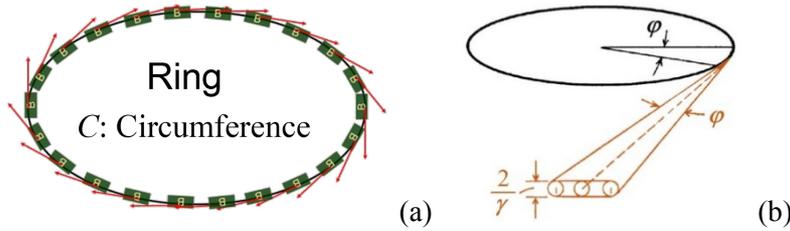

**Fig. 4:** (a) Schematic configuration of a ring with a circumference of C, where "B" means a dipole magnet, and (b) synchrotron radiation emitted within a deflection angle of $\varphi$.

The average power line density in the ring is obtained by

$$\langle P_{Ie,line} \rangle = P_{Ie} / C. \quad (17)$$

The power in an angle of $\varphi$ is

$$P_{Ie}(\phi) = P_{Ie} \frac{\phi}{2\pi}. \quad (18)$$

Until now, we have only considered SR in a time domain regime. Sometimes it is also useful and important to consider SR in the frequency domain. The frequency spectrum of the electric field is obtained by Fourier transform of $E(t)$:

$$\tilde{E}(\omega) = \frac{1}{\sqrt{2\pi}} \int_{-\infty}^{+\infty} E(t) e^{i\omega t} dt \quad . \tag{19}$$

From Eq. (5),

$$\frac{dP}{d\Omega} = \vec{n} \cdot \vec{S} R^2 \Big|_{\text{ret}} = \varepsilon_0 c E^2 R^2 \Big|_{\text{ret}} \quad , \tag{20}$$

$$\frac{dW}{d\Omega} = \int \frac{dP(t)}{d\Omega} dt = \frac{1}{\mu_0 c} \int_{-\infty}^{+\infty} (RE)^2 \, dt = \frac{1}{\mu_0 c} \int_{-\infty}^{+\infty} \left| R\tilde{E}(\omega) \right|^2 d\omega \quad . \tag{21}$$

Then, the frequency spectrum of power is given by

$$\frac{d^2 W}{d\Omega d\omega} = \frac{1}{\mu_0 c} \left( R\tilde{E}(\omega) \right)^2 = \frac{1}{2\pi\mu_0 c} \left| \int_{-\infty}^{+\infty} (RE) e^{i\omega t} dt \right|^2$$

$$= \frac{e^2}{16\pi^3 \varepsilon_0 c} \left| \int_{-\infty}^{+\infty} \left[ \frac{\left| \vec{n} \times \left\{ (\vec{n} - \vec{\beta}) \times \dot{\vec{\beta}} \right\} \right|^2}{(1 - \vec{n} \cdot \vec{\beta})^5} \right]_{\text{ret}} e^{i\omega\left(t' + \frac{R(t')}{c}\right)} dt' \right|^2 \quad . \tag{22}$$

Finally, the spatial and spectral energy distribution of SR per unit frequency and solid angle is given by

$$\frac{d^2 W}{d\Omega d\omega} = \frac{e^2}{16\pi^3 \varepsilon_0 c} \gamma^2 \frac{\omega^2}{\omega_c^2} K_{2/3}^2(\xi) F(\xi, \theta) \quad , \tag{23}$$

$$\xi \equiv \frac{1}{2} \frac{\omega}{\omega_c} \left( 1 + \gamma^2 \theta^2 \right)^{3/2} \quad , \tag{24}$$

$$F(\xi, \theta) \equiv \left( 1 + \gamma^2 \theta^2 \right)^2 \left[ 1 + \frac{\gamma^2 \theta^2}{1 + \gamma^2 \theta^2} \frac{K_{1/3}^2(\xi)}{K_{2/3}^2(\xi)} \right] \quad . \tag{25}$$

where $K_i(\zeta)$ is the modified Bessel function. The former term in Eq. (25) is the $\sigma$ mode, where the electrical field is orthogonal to the deflecting field ($B$). The latter term is the $\pi$ mode, where the electrical field is in the plane of the deflecting field and the line of observation. For high energies, the $\sigma$ mode is dominant. Here,

$$\omega_c \equiv \frac{3}{2} \frac{c\gamma^3}{\rho} \tag{26}$$

is the critical frequency. This is the frequency that halves the total energy.

The photon number (photon flux) with a beam current $I_e$ per unit solid angle and frequency is given by

$$\frac{d^2 \dot{N}_{\text{ph,Ie}}}{d\Omega (d\omega/\omega)} = \frac{d^2 P_{\text{Ie}}}{d\Omega d\omega} \frac{1}{\hbar} = \frac{d^2 W}{d\Omega d\omega} \frac{I_e}{e} \frac{1}{\hbar} \quad , \tag{27}$$

where $\hbar \equiv h/2\pi$. The spatial and spectral photon flux distribution per unit solid angle $d\theta d\psi$ and the band width $d\omega/\omega$ that is, the brightness, is given by

$$\frac{d^3 \dot{N}_{ph,Ie}}{d\theta d\psi (d\omega/\omega)} = C_\omega E^2 I_e \frac{\omega^2}{\omega_c^2} K_{2/3}^2(\xi) F(\xi,\theta) \ , \tag{28}$$

$$C_\omega \equiv \frac{3\alpha}{4\pi^2 e(m_e c^2)^2} = 1.3255 \times 10^{22} \frac{\text{photons}}{\text{s rad}^2 \text{ GeV}^2 \text{ A}}$$
$$= 1.3255 \times 10^{13} \frac{\text{photons}}{\text{s mrad}^2 \text{ GeV}^2 \text{ A 0.1\% bandwidth}} \ , \tag{29}$$

where $\alpha$ is the fine-structure constant. The brightness is a key parameter in considering the performance of light (photon) sources. Figure 5 shows the typical brightness as a function of photon energies and the critical energies.

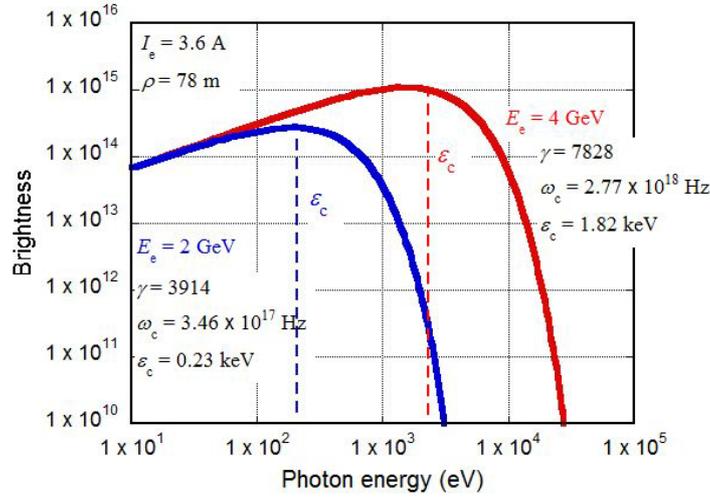

**Fig. 5:** Examples of brightness. $\varepsilon_c$ is also indicated

The critical energy of photons is given by:
$$\varepsilon_c = \frac{3}{2} \frac{\hbar c \gamma^3}{\rho} \equiv \hbar \omega_c \ , \tag{30}$$

$$\varepsilon_c \ [\text{eV}] = 2218 \times 10^3 \times \frac{E_e \ [\text{GeV}]^3}{\rho \ [\text{m}]} = 0.665 \times 10^3 \times E_e \ [\text{GeV}]^2 B \ [\text{T}] \ . \tag{31}$$

Using the critical energy, the mean photon energy is expressed as
$$\langle \varepsilon \rangle = \frac{8}{15\sqrt{3}} \varepsilon_c \ , \tag{32}$$

and the total photon flux is
$$\dot{N}_{ph} = \frac{15\sqrt{3}}{8} \frac{P_{tot}}{\varepsilon_c} . \tag{33}$$

The critical energy is a key parameter characterizing SR.

The total photon number is given by integrating Eq. (28) with respect to $\theta$, $\psi$ (that is, the whole ring), and $\omega$,

$$\dot{N}_{\text{ph,Ie}} = \frac{15\sqrt{3}}{4} C_{\psi} I_e E_e = 8.08 \times 10^{20} I_e \text{ [A]} E_e \text{ [GeV]} \text{ [photons s}^{-1}\text{]} , \qquad (34)$$

$$C_{\psi} \equiv \frac{4\alpha}{9 e m_e c^2} = 3.9614 \times 10^{19} \frac{\text{photons}}{\text{s rad GeV A}} . \qquad (35)$$

The average photon number per unit length along the ring is obtained by
$$\left\langle \dot{N}_{\text{ph,Ie,line}} \right\rangle = \dot{N}_{\text{ph,Ie}} / C . \qquad (36)$$

The photon number for an angle of $\varphi$ is
$$\dot{N}_{\text{ph,Ie}}(\phi) = \dot{N}_{\text{ph,Ie}} \frac{\phi}{2\pi}. \qquad (37)$$

## 3  Effects of synchrotron radiation

The three main effects of SR on an accelerator (especially on the vacuum system) (see Fig. 6) are given below.

i) Heat load: when SR hits a surface, it transfers the energy to the surface. The SR heats up the beam pipe, and sometimes damages it by excess heating and thermal stress.

ii) Gas load: when SR hits a surface, it desorbs gas molecules on the surface. The gas desorption increases vacuum pressure. The pressure rise reduces the beam lifetime and increases background noise in the detector.

iii) Emission of electrons: when SR hits a surface, the surface emits electrons (photoelectrons). The emitted photoelectrons enhance the formation of the electron cloud, which leads to electron cloud instabilities (for positive beams).

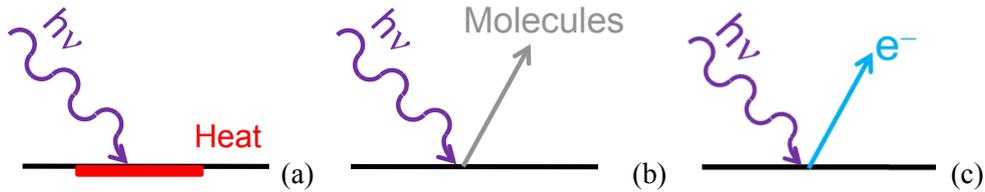

**Fig. 6:** Three major effects of SR. (a) heat load; (b) gas load; (c) electron emission

### 3.1  Heat load

#### 3.1.1  General

When SR hits the inner wall of a beam pipe, it transfers energy to the surface, resulting in heating of the surface. As described above, the SR 'beams' (concentrates) in the front direction. If the irradiated area is not properly cooled, the surface is easily excessively heated and damaged. For example, if $E_e = 4$ GeV, $I_e = 3.6$ A, $\rho = 74$ m, and $C = 2000$ m (parameters for SuperKEKB [2]), from Eq. (16),

$$\left\langle P_{\text{Ie,line}} \right\rangle = 88.5 \times 10^3 \times 4^4 \times 2.6 / 74 / 2000 = 550 \text{ W m}^{-1} . \qquad (38)$$

The power density is sufficiently high to melt metals if the irradiation area is not cooled. The heat load is actually distributed along the ring. The sources (emitting points) are in the bending magnets. For a uniform beam pipe, the heat load is maximum in the bending magnet, and decreases gradually on the downstream side, as shown in Fig. 7 for the SuperKEKB [2]. In this case the average power line density is ~0.6 kW m$^{-1}$, but the peak power line density is 2.3 kW m$^{-1}$. When considering heating by SR the maximum power density is more important than the average.

Dependencies of the SR power line and area densities on the distance from the emitting point to the irradiated point $R$ and the incident angle $\theta_i$ are shown in Fig. 8. The line power density in the magnet is proportional to $1/R$, and is almost constant, because $R$ and $\theta_i$ are constant. The SR power line density outside the magnet, on the other hand, is proportional to $1/R \times \theta_i \propto 1/R^2$, because the incident angle $\theta_i$ is also almost proportional to $1/R$. For the power area density, on the other hand, the vertical spread angle of $2/\gamma$ should be taken into account. Power area density is key in evaluating the thermal stress.

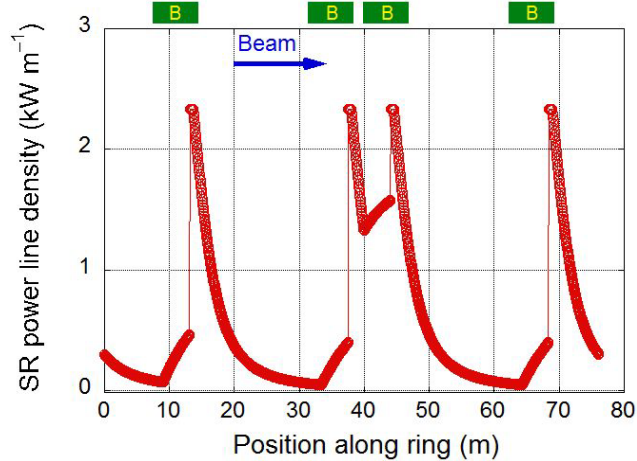

**Fig. 7:** Example of the distribution of SR power line density (SuperKEKB [2])

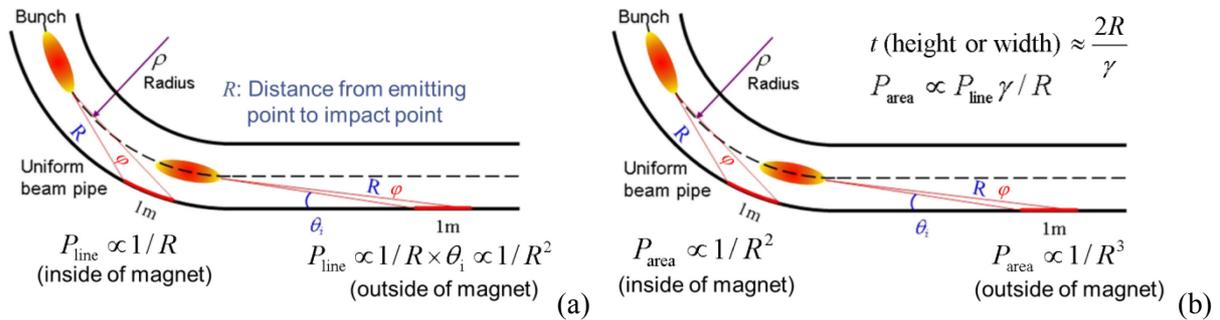

**Fig. 8:** Typical dependencies of SR power (a) line and (b) area densities upon $R$ and $\theta_i$

### 3.1.2 Countermeasures

The basic principle of the heat load countermeasure is to receive SR at specific places (photon stops) with a proper cooling system at large $R$ and small $\theta_i$. In general, this can be achieved in two ways: using distributed photon stops (photon masks) and localized photon stops.

#### 3.1.2.1 Distributed photon stops

In this case, small photon stops are placed upstream of bellows chambers or flanges to make short SR shadows, as shown in Figs. 9 and 10 [3–6]. The photon stops have a relatively low height ($H$) of ~10 mm. The typical shadow length, given by $H/\tan \theta_i$, is 200–400 mm. Most of the heat load is distributed along the ring, and the heat load at the photon stops is relatively small ($\theta_i$ is also small). The structure of the beam pipes is simple. Distributed photon stops must be used if the power density at the localized photon stop (described below) is too high.

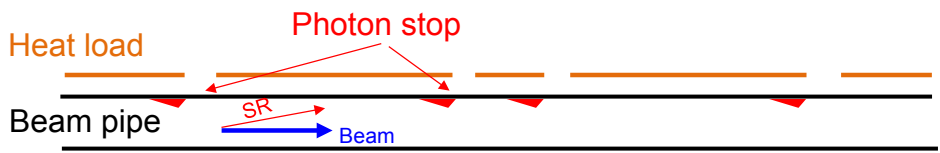

**Fig. 9:** Schematic configuration of distributed photon stops

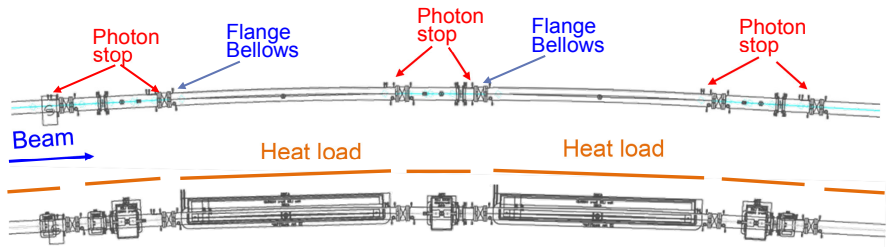

**Fig. 10:** Example of the distributed photon stop scheme (SuperKEKB [2])

*3.1.2.2 Localized photon stops*

In this case, large photon stops that result in long SR shadows are placed locally, as shown in Figs. 11 and 12 [7–10]. The typical mask height is 100–200 mm, and the typical shadow length is a few metres of SR, i.e. the photon stops receive the SR power corresponding to the power of a few metres. This means that most of the heat load concentrates in the photon stops, usually at a much higher power density than in the case of the distributed photon stops. One of the criteria used to decide on a particular photon stop scheme, distributed or localized, is the SR power density. Sometimes, in light sources, the photon stop is called a 'crotch absorber'. The structure of the beam pipe is likely to be complicated here. Effective pumping is realized by putting pumps at the same places as the photon stops (see below).

Various types of photon stops (masks) have been designed in various accelerators [7–12]. In designing the photon stops, simulations using finite element methods (FEM) are very useful in evaluating the temperature and stress distribution. Key design points are:

i) to obtain a slanting irradiated surface (i.e. make $\theta_i$ as small as possible) to reduce the power density in the horizontal direction as well as in the vertical direction;

ii) to prepare sufficient and effective cooling paths;

iii) to use materials with high thermal conductivity and high thermal strength.

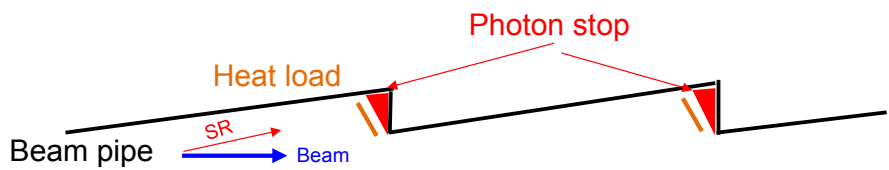

**Fig. 11:** Schematic configuration of the localized photon stops

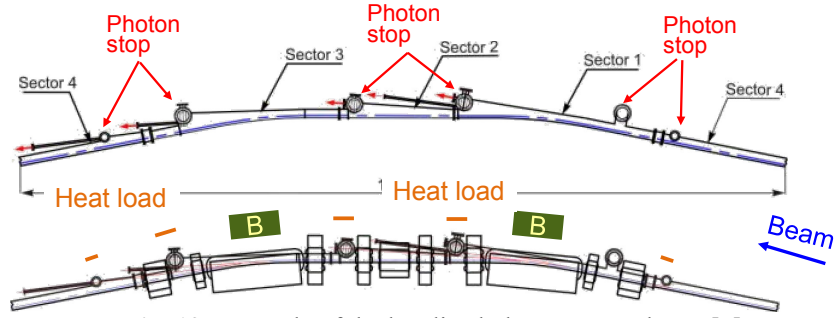

**Fig. 12:** Example of the localized photon stop scheme [7]

#### 3.1.2.3 Other countermeasures

i) Use materials with a high thermal conductivity and a high thermal strength, e.g. copper, copper–chromium alloys, and GlidCop [13].

ii) Use beam pipes with an antechamber, where the SR hits a point far from the emission point. The power area density decreases as $R$ increases.

### 3.2 Gas load

#### 3.2.1 General

When SR hit the inner surface of the beam pipe, it desorbs gas molecules adsorbed on the surface. This is called photon stimulated gas desorption (PSD) [9, 14–16]. Residual gases in beam pipes during the operation are the result of PSD. For example, if $E_e$ = 4 GeV and $\rho$ = 74 m, the critical energy $\varepsilon_c$ is 1.9 keV. Because the temperature corresponding to 1 eV is approximately 12 000°C, a 1 keV photon is enough to destroy the chemical bonding between adsorbed molecules and surface molecules (a few electron volts). PSD is also much more effective than baking. Considerable gas desorption is expected with PSD compared to with thermal gas desorption for large photon numbers.

The main effects of the gas load are given below.

i) Energy loss due to scattering with the residual gases. This leads to particle loss and then a shorter beam lifetime. The loss of particles increases the background noise of the detectors, and can also cause radio activation.

ii) Generation of ions by ionization of residual gases. Ion instabilities can be excited in beams with negative charges, such as electrons. Furthermore, ions generated by ionization of residual gases hit the beam pipe wall and desorb gases from the surface (this is called ion stimulated gas desorption (ISD)). This sometimes leads to a pressure instability.

Here, we briefly discuss the beam lifetime $\tau$, which is defined as

$$I_e = I_{e0} e^{-\frac{t}{\tau}} \ . \tag{39}$$

where $I_e$ and $I_{e0}$ are the beam current and the initial beam current, respectively. For sufficiently large apertures, $\tau$ can be usually expressed by

$$\frac{1}{\tau} = \sum_i \{\sigma_B(Z_i) + Z_i \sigma_M + \sigma_R(Z_i)\} p_i \ . \tag{40}$$

Here, $\sigma_B$, $\sigma_M$ and $\sigma_R$ are the cross-sections of three major interaction processes with gas molecules, i.e. a Bremsstrahlung with nuclei, Moller scattering with electrons outside nuclei, and Rutherford scattering with nuclei, respectively. As indicated by this equation, lifetime is inversely proportional to the pressure.

### 3.2.2 Photon stimulated gas desorption

The SR irradiation on the inner surface results in the emission of photoelectrons. The photon energy is high enough to cause electron emission (photoelectrons) from material surfaces where the work functions are a few electron volts. The quantum efficiency $\eta_e$, the yield of photoelectrons per photon, is ~0.1 electrons photon$^{-1}$. We will discuss electron emission from the surface in the section below. These electrons hitting the surface desorb molecules from the surface, because they also have sufficiently high energies. This is called electron stimulated gas desorption (ESD). It is believed that most PSD comes from ESD.

The number of gas molecules emitted by one photon is the PSD rate. It is usually expressed by $\eta$ [molecules photon$^{-1}$]. After the usual baking, the major desorbed gases are hydrogen ($H_2$), carbon monoxide (CO), and carbon dioxide ($CO_2$) (Fig. 13) [9]. Water ($H_2O$) is the main gas for a non-baked system.

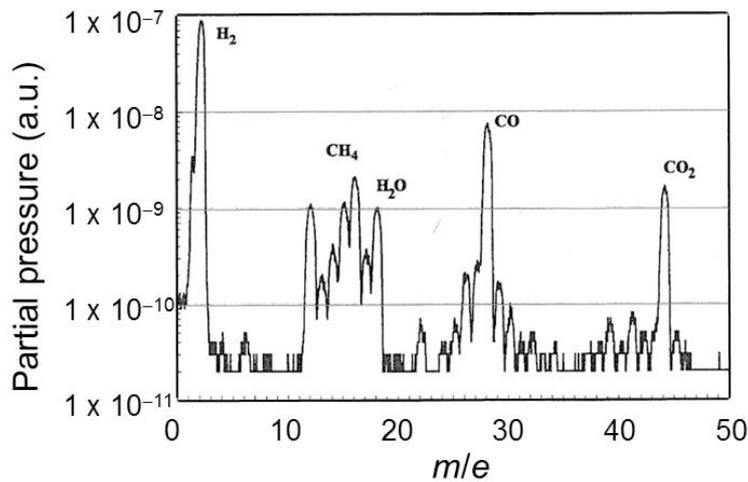

**Fig. 13:** Typical mass spectrum of the residual gases due to PSD [9]

The $\eta$ increases with the incident photon energy (critical energy) because the deposit energy increases, as shown in Fig. 14 [14].

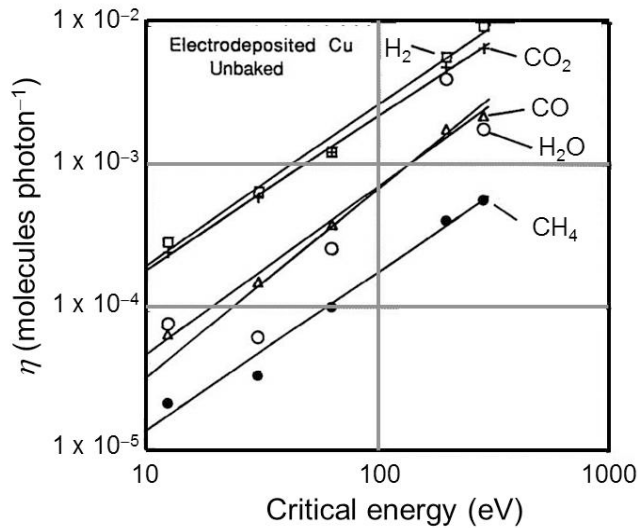

**Fig. 14:** Dependence of $\eta$ on the critical energy of SR [14]

The $\eta$ increases as the incident angle decreases [15]. A rough surface, therefore, can result in a decrease in $\eta$. Note that if the surface is smooth and the incident angle is small, the reflection of SR should be taken into account.

Another important property of PSD is aging (scrubbing). The $\eta$ decreases with the integrated photon number (photon dose, $D$), as shown in Fig. 15. This phenomenon is called beam aging or scrubbing. Typical values of $\eta$ before SR irradiation are $10^{-3}$ to $10^{-2}$ molecules photon$^{-1}$. The $\eta$ decreases to $10^{-6}$ to $10^{-7}$ after sufficient aging. Usually, $\eta$ varies according to the function

$$\eta = D^{-1 \sim -0.6} . \tag{41}$$

Practically, when designing the vacuum system, an $\eta$ of $1 \times 10^{-5}$ to $1 \times 10^{-6}$ molecules photon$^{-1}$ is assumed considering the aging effect. $\eta$ also strongly depends on the surface condition, the degree of contamination, the thickness of the oxide layer, the roughness of the surface, etc. [16].

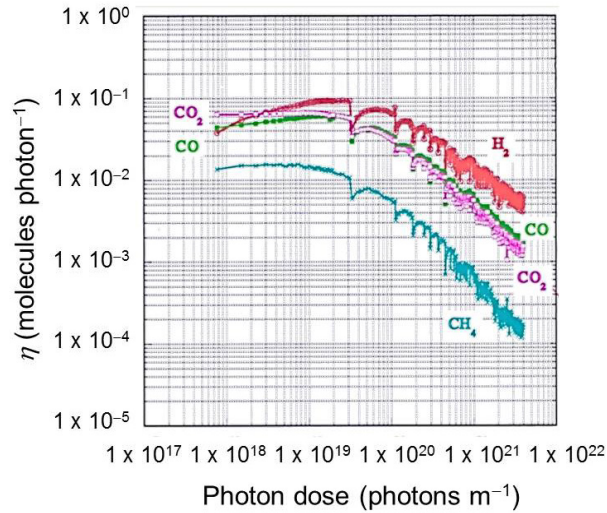

**Fig. 15:** Typical beam scrubbing (aging) of $\eta$

Here we estimate the gas load. For example, if $E_e$ = 4 GeV, $I_e$ = 2.6 A and $C$ = 3000 m, from Eq. (36),

$$\left\langle \dot{N}_{\text{ph,Ie,line}} \right\rangle = 8.08 \times 10^{20} \times 4 \times 2.6 / 3000 = 2.8 \times 10^{18} \text{ photons s}^{-1}\text{m}^{-1} . \tag{42}$$

If $\eta = 1 \times 10^{-6}$ molecules photon$^{-1}$,

$$\left\langle \dot{N}_{\text{ph,Ie,line}} \right\rangle = 2.8 \times 10^{18} \times 1 \times 10^{-6} = 2.8 \times 10^{12} \text{ molecules s}^{-1} \text{ m}^{-1} . \tag{43}$$

The average line gas desorption rate (gas load) along the ring $Q_{\text{av,line}}$ (at $T$ = 25°C = 298 K) is

$$\begin{aligned} Q_{\text{av,line}} &= \left\langle \dot{N}_{\text{mol,Ie,line}} \right\rangle \times k_B T = 2.8 \times 10^{12} \times 1.38 \times 10^{-23} \times 298 \\ &= 1.1 \times 10^{-8} \text{ Pa m}^3 \text{ s}^{-1} \text{ m}^{-1} . \end{aligned} \tag{44}$$

Here we use the ideal gas law equation

$$pV = N_{\text{mol}} k_B T . \tag{45}$$

Here, $V$ and $k_B$ are the volume and Boltzmann constant, respectively. The expression is convenient in designing a vacuum system, because the pumping speed is usually expressed in cubic metres per second (m$^3$ s$^{-1}$). If the average linear pumping speed is $S_{\text{av,line}}$ (m$^3$ s$^{-1}$ m$^{-1}$) in the ring, the obtained average pressure $p_{\text{av}}$ (Pa) is

$$p_{av} = \frac{Q_{av,line}}{S_{av,line}} \quad \text{Pa} \quad . \tag{46}$$

The distribution of gas load is almost the same as that of photons. Basically, the gas load is high downstream of the bending magnets, as in the case of heat load (Fig. 16). In this case, the average photon line density is ~5.5 × $10^{18}$ photons $s^{-1}$ $m^{-1}$. The maximum photon line density is ~3 × $10^{19}$ photons $s^{-1}$ $m^{-1}$.

Note here that the distribution of gas load is not exactly the same as for photons, because the PSD depends on the beam dose and $\theta^i$. Actually, the difference between the maximum and the minimum values decreases with time.

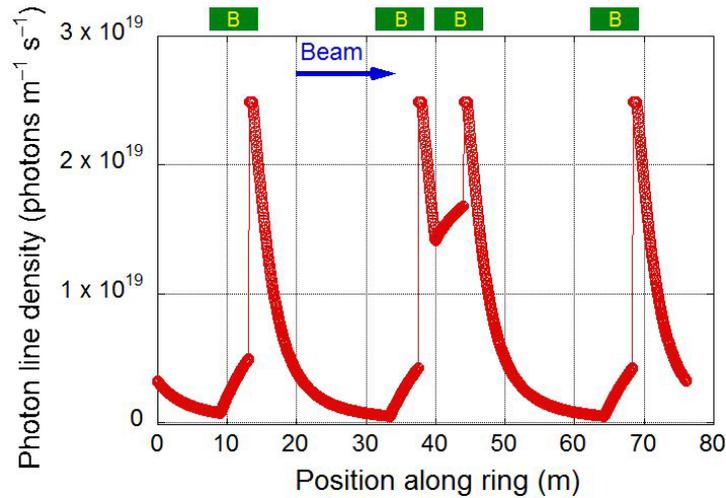

**Fig. 16:** Distribution of photon line density along the ring (SuperKEKB) [2]

### *3.2.3 Countermeasures*

The basic principle of countermeasures is to prepare proper pumps at places where photons irradiate the beam pipe. There are two ways to treat the gas load: distributed pumping and localized pumping.

#### *3.2.3.1 Distributed pumping*

Distributed pumping works well with the distributed photon stops described above, as shown in Fig. 17 [3–6]. Beam pipes are usually very narrow and long; hence their conductance is small, typically <0.1 $m^3$ $s^{-1}$ $m^{-1}$. In the distributed pumping scheme, pumps are located along the beam pipe, just alongside the beam channel, and the beam pipe is then effectively evacuated. Uniform pumping speed along the ring is realized. Distributed pumping is effective where the gas loads are distributed evenly along the beam pipe. In a distributed pumping scheme, the structure of the beam pipes is relatively simple.

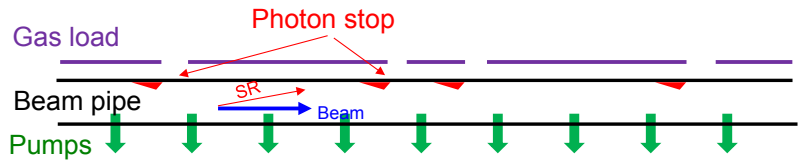

**Fig. 17:** Distributed pumping scheme

A commonly used pump in the distributed pumping scheme is the distributed sputter-ion pump (DIP). This is a sputter-ion pump operating in the bending magnets' magnetic field. The DIP was popular until ca. 1990. Another commonly used pump is the non-evaporable getter (NEG) pump. The

NEG strips are placed along the beam pipe. A thin film of the NEG ingredient coating inside the beam pipe has recently been used in various facilities. In the case of the previous example, if we use a distributed pumping system with an average pumping speed of ~0.11 m$^3$ s$^{-1}$ m$^{-1}$, an average pressure of 2.3 × 10$^{-7}$ Pa is obtained as shown in Fig. 18, assuming $\eta = 1 \times 10^{-6}$ molecules photon$^{-1}$. A similar pressure profile to that of the photon line density is obtained.

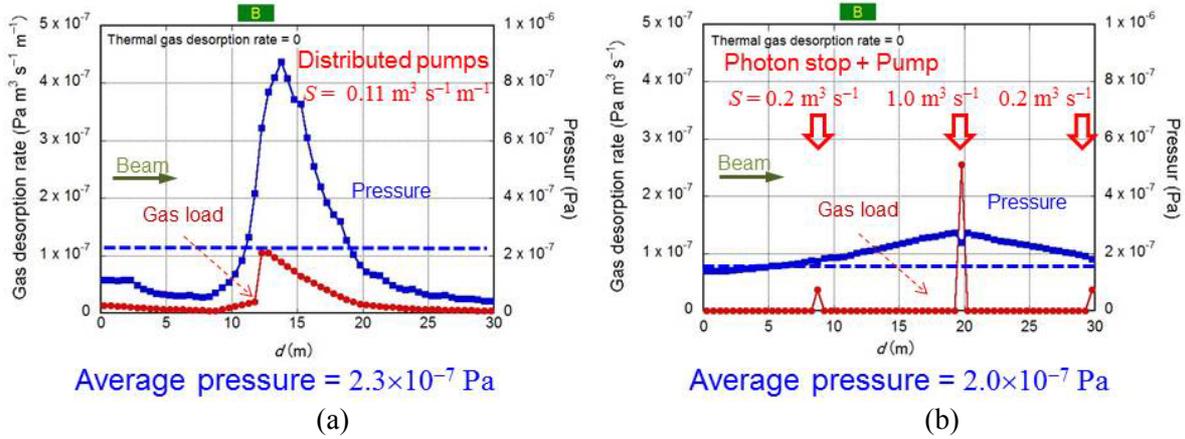

(a)  (b)

**Fig. 18:** Comparison of distributed and localized pumping scheme, where the photon line density in Fig. 15 and $\eta = 1 \times 10^{-6}$ molecules photon$^{-1}$ are assumed. (a) Distributed pumping; (b) localized Pumping.

### 3.2.3.2 *Localized pumping*

The localized pumping scheme works well with the localized photon stops, as shown in Fig. 19 [7–12]. In this case, the pumps are placed near the localized photon stops, usually downstream of the bending magnets. The localized photon scheme is related to a localized gas load. The pumps are concentrated at the locations where the gas loads are large. This is a reasonable approach to achieve ultra-high vacuum, and is adopted in many recent photon sources. The widely used pumps are turbo-molecular pumps, sputter ion pumps, Ti-sublimation pumps, and NEG cartridges, etc. Here the structure of the beam pipes is more than in the case of the distributed pumping scheme. As indicated in Fig. 18, if localized pumps are used and the thermal gas desorption is ignored, a lower average pressure is obtained compared to that obtained with distributed pumping, even with smaller total pumping speeds. Note, however, that low thermal gas desorption is essential. Otherwise the pressure between the adjacent pumps will be high owing to the limited conductance of the beam pipe.

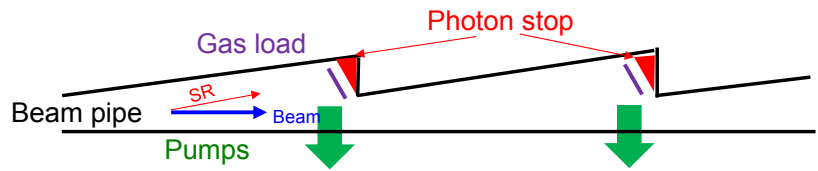

**Fig. 19:** Localized pumping scheme

### 3.2.3.3 *Other countermeasures*

An important measure is to avoid contamination during the beam pipe's manufacturing and assembly process. A clean assembly environment should be ensured. Surface treatments, such as chemical cleaning, argon glow discharge, and pre-baking, are effective in reducing thermal gas desorption. Using an antechamber scheme is also effective, because photons hit photon stops in the antechamber, which is separated from the beam channel. Desorbed gas is confined within the antechamber. The antechamber structure is usually adopted for a localized photon stop scheme. The antechamber scheme also provides a relatively smooth beam channel, which contributes to a lower beam impedance.

## 3.3 Electron emission

### 3.3.1 General

When SR hits the surface, photoelectrons are emitted from the surface, as described above. The yield of photoelectrons from one incident photon, the quantum efficiency $\eta_e$, is ~0.1 electrons photon$^{-1}$. If the beams are positively charged (i.e. positrons or protons), they attract the electrons. The electrons that are accelerated by the next bunch's electric fields hit the surface and emit electrons, which are called secondary electrons. If the secondary electron yield (SEY), the number of electrons emitted by one incident electron, is larger than 1, the enhancement of electrons (multipactoring) can occur. This positive feedback leads to the accumulation of electrons around the beams. This group of electrons is called the electron cloud [17].

### 3.3.2 Secondary electron yield

Figure 20 shows the creation process and the typical energy spectrum of secondary electrons [18]. The secondary electrons are emitted from the surface following the cosine law, i.e., uniformly. The energy of secondary electrons is less than 50 eV. The SEY depends on the incident angle of electrons as in the case of the photoelectron yield's dependence on the incident angle of the photons. The SEY $\delta$ increases with the incident angle $\theta$, which is the angle between the direction of the incident electron and the normal to the surface [19]. The dependence can be explained as follows: for shallow incidence (large $\theta$), electrons generated along the path of the incident electron can easily escape to vacuum (see Fig. 20). The following two formulae are commonly used in simulations. For $\theta \sim 0°$,

$$\delta \approx \frac{\delta_0}{\cos\theta} \quad (47)$$

Here $\delta_0$ is the SEY at normal incidence of the primary electron. For $\theta \to 90°$,

$$\delta \approx \delta_0 e^{\alpha X_m (1-\cos\theta)} \quad (48)$$

Here, $X_m$ is the depth at which secondary electrons are generated at normal incidence and $\alpha$ is the absorption rate. Usually, $\alpha X_m \sim 0.4$ is used.

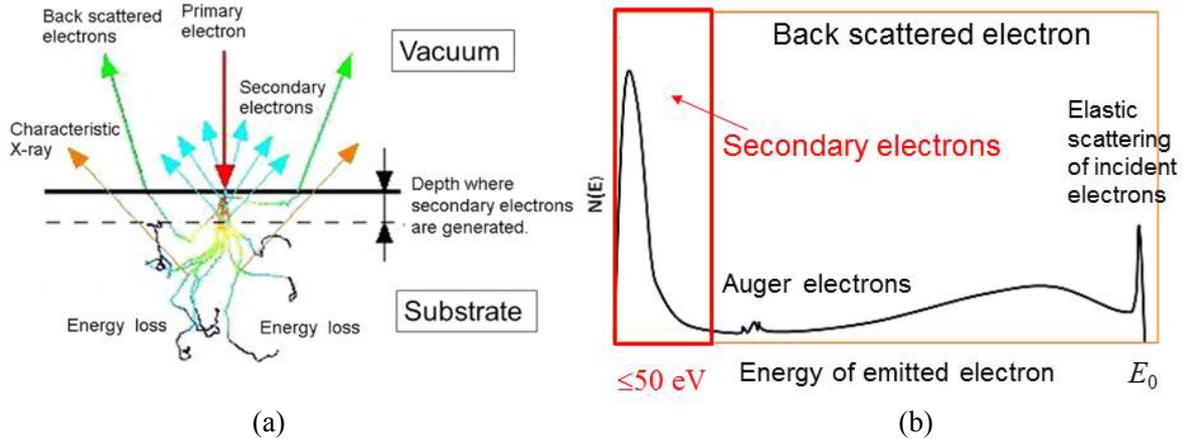

**Fig. 20:** (a) Process of generating secondary electrons; (b) typical energy spectrum of secondary electrons [18]

Figure 21 shows the dependence of SEY on the energy of the incident electron (primary electron). The SEY has a maximum at an incident electron energy of 200–400 eV and decreases gradually with increasing energy. Two formulae for $\delta$ are usually used for the simulation [20–24]. One of these is

$$\delta(E_r) \approx \delta_{\max} 1.11 E_r^{-0.35} \left(1 - e^{-2.3 E_r^{1.35}}\right), \quad (49)$$

where $\delta_{max}$ is the maximum yield for perpendicular incident, $E_r \equiv E_p/E_{pm}$, where $E_p$ is the energy of the incident electron, and $E_{pm}$ is the primary electron energy at which the yield is at a maximum.

The other expression is:

$$\delta(E_r) \approx \delta_{max} \frac{sE_r}{s-1+E_r^s} \qquad (50)$$

where $s \sim 1.4$.

A decrease in SEY with electron dose (integrated electrons per unit area) is observed, as in the case for the PSD rate ($\eta$) [25]. The decrease is also called as the aging or conditioning. The SEY also strongly depends on the surface conditions and materials.

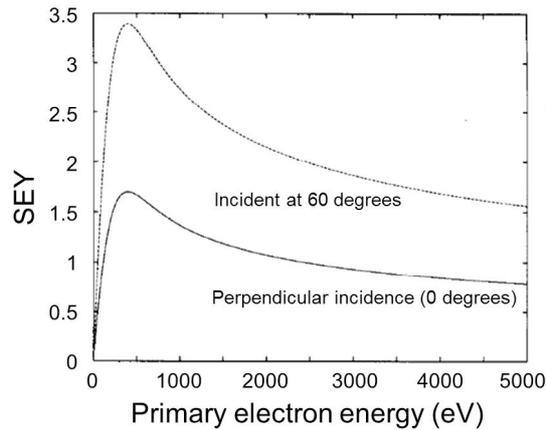

**Fig. 21:** Dependence of SEY on the energy of the incident electrons [20]

### 3.3.3 Electron cloud instability

This section briefly introduces the electron cloud effect (ECE) [17, 26–28]. If the electron density around the beam exceeds a threshold value, the electron cloud causes a beam instability, called the electron cloud instability. First, the displacement of the top bunch affects the following bunches via the electron cloud. Then, the perturbation of the electron cloud (a kind of wake field) affects the following bunches or the electrons in the same bunch. The former is called a coupled bunch instability, and the latter is called a head–tail instability, as explained in Fig. 22. The electron cloud instability leads to the blow-up of beam size, which increases the emittance of the beam, and decreases in the luminosity in the colliders. The electron cloud instability is a critical issue in recent high-intensity proton and positron storage rings. Many theoretical and experimental studies have been conducted about the formation of the electron cloud, the simulation of beam instability, the measurement of beam emittance, and countermeasures.

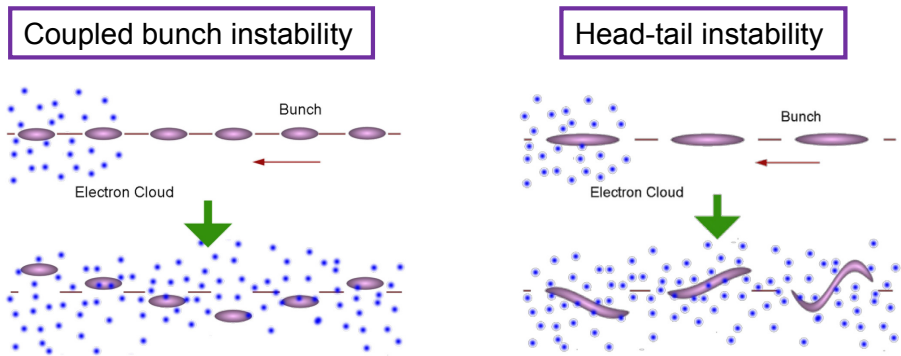

**Fig. 22:** Two types of electron cloud instabilities

Here we roughly estimate the number of generated photoelectrons. For the SuperKEKB positron ring, where $E_e$ = 4 GeV, $I_e$ = 3.6 A and $C$ = 3000 m, the average photon linear density along the ring from Eq. (36) is

$$\langle \dot{N}_{\text{ph,Ie,line}} \rangle = 8.08 \times 10^{20} \times 4 \times 3.6 / 3000 = 3.9 \times 10^{18} \text{ photons s}^{-1} \text{ m}^{-1} . \tag{51}$$

If the quantum efficiency ($\eta_e$) is 0.1, the number of emitted photoelectrons is

$$\langle \dot{N}_{\text{ele,Ie,line}} \rangle = \eta_e \times \langle \dot{N}_{\text{ph,Ie,line}} \rangle = 3.9 \times 10^{17} \text{ electrons s}^{-1}\text{m}^{-1} . \tag{52}$$

In the case of the SuperKEKB, the threshold of the electron density, $\rho_{e,\text{th}}$, to result in the head–tail instability is $2 \times 10^{11}$ electrons m$^{-3}$. It has been found that $\rho_{e,\text{th}}$ is easily achieved if no countermeasures are adopted.

### 3.3.4 *Countermeasures*

The basic principles of the countermeasures are to suppress electron emissions and remove electrons around the beams. Various countermeasures have been proposed and studied, and some have been applied in practice.

#### 3.3.4.1 *Beam pipe with antechambers*

The SR irradiates the side wall of the antechamber, far from the beam (Fig. 23) [3, 29]. Hence, the photoelectrons do not easily interface with the beam. Note that some photons may hit the outside of the antechamber at points far from the photon emitting point owing to the vertical spread of ~2/$\gamma$. Furthermore, multipactoring of secondary electrons is more significant for a large beam current. The antechamber structure is, therefore, effective at low beam currents.

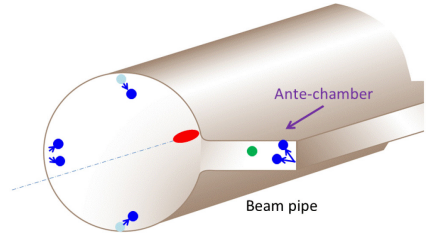

**Fig. 23:** Beam pipe antechamber

#### 3.3.4.2 *Inner coating with a low SEY*

At high beam currents, the main mechanism forming the electron cloud is the multipactoring of secondary electrons (Fig. 24) [30–34]. In this situation, some inner coatings with a low SEY are effective in suppressing the electron cloud formation. Possible candidates are TiN, graphite, and NEG ingredients.

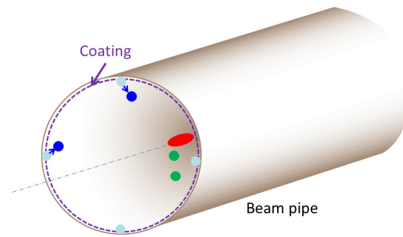

**Fig. 24:** Beam pipe inner coating

### 3.3.4.3 Grooved surface

A surface with a grooved structure is found to have a low SEY (Fig. 25) [35–37]. The SEY is structurally reduced, especially in a magnetic field. A coating of material with a low SEY on the groove enhances the reduction of SEY. Beam impedance may be a concern.

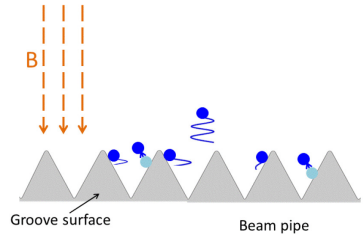

**Fig. 25:** Grooved surface

### 3.3.4.4 Solenoid field

With a magnetic field along the beam pipe, the electrons emitted from the surface return to the surface due to Larmor motion (Fig. 26) [38, 39]. Emitted photoelectrons or secondary electrons have an energy of several tens of electron volts. Hence, a magnetic field of several tens of gauss is sufficient. Drastic effects were observed in PEP-II and KEKB.

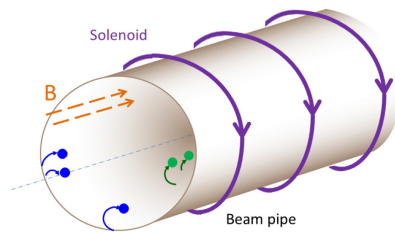

**Fig. 26:** Solenoid field

### 3.3.4.5 Clearing electrode

An electrode in a beam pipe with a high positive potential attracts the electrons around the beam orbit (Fig. 27) [23, 37, 40, 41]. A drastic effect in reducing electron density is expected and has been confirmed in experiments. The effect was also demonstrated at DAFNE. Beam impedance is also a concern with this countermeasure.

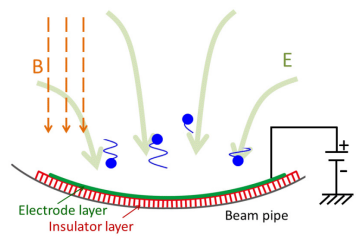

**Fig. 27:** Clearing electrode

### 3.3.4.6 Other countermeasures

Alternative effective countermeasures are making a wide bunch gap and changing the bunch filling patterns. Experience with KEKB, however, suggests that these techniques are not such effective countermeasures.